\documentclass[aps,prb,showpacs,twocolumn,groupedaddress,floatfix]{revtex4}
\usepackage{subfigure}
\usepackage{amsmath}
\usepackage{graphicx}
\usepackage{rotating}
\usepackage{amsfonts}
\usepackage{amssymb}

\begin{document}

\title{Sr$_{2}$VO$_{3}$FeAs as Compared with Other Fe-based Superconductors}
\author{I.I. Mazin}
\affiliation{Code 6393, Naval Research Laboratory, Washington, D.C. 20375}
\date{Printed on \today}

\begin{abstract}
One of the most popular scenarios for the superconductivity in Fe-based
superconductors (FeBSC) posits that the bosons responsible for electronic
pairing are spin-fluctuations with a wave vector spanning the hole Fermi
surfaces (FSs) near $\Gamma$ and the electron FSs near M points. So far all
FeBSC for which neutron data are available do demonstrate such excitations,
and the band structure calculations so far were finding quasi-nested FSs in
all FeBSC, providing for a peak in the spin susceptibility at the desired
wave vectors. However, the newest addition to the family, Sr$_{2}$VO$_{3}$%
FeAs, has been calculated to have a very complex FS with no visible
quasi-nesting features. It was argued therefore that this material does not
fall under the existing paradigm and calls for revisiting our current ideas
about what is the likely cause of superconductivity in FeBSC. In this paper,
I show that the visible complexity of the FS is entirely due to the
V-derived electronic states. Assuming that superconductivity in Sr$_{2}$VO$%
_{3}$FeAs, as in the other FeBSC, originates in the FeAs layers, and the
superconducting electrons are sensitive to the susceptibility of the FeAs
electronic subsystem, I recalculate the bare susceptibility, weighting the
electronic states with their Fe character, and obtain a susceptibility that
fully supports the existing quasi-nesting model. 
Besides, I find that the mean-filed magnetic ground state is the checkerboard
in V sublattuce and stripes in the Fe sublattice.
\end{abstract}

\maketitle

The recently discovered\cite{disc} Fe-based high-temperature superconductors (FeBSC)
represent a challenging case for the theory of superconductivity. They
appear to be rather different from cuprates in terms of their electronic
structure, magnetic order, correlation effects, and superconducting symmetry%
\cite{review}. So far the most popular suggestion for the pairing mechanism
has been one that assigns the role of an intermediate boson to spin
fluctuations with wave vectors close to \textbf{Q=(}$\pi,\pi)$ (in the
two-Fe Brillouin zone). There are two ways to generate such spin
fluctuations: one assumes superexchange between the second neighbors in the
Fe lattice and the other exploits the fact that the non-interacting spin
susceptibility calculated using the one-electron band structure has a peak,
or better to say a broad maximum close to \textbf{(}$\pi,\pi)$ (see review
Ref. \onlinecite{review}). A strong argument in favor of the latter scenario
was the case of FeSe, where the parent magnetic compound FeTe shows an
antiferromagnetic order at a different wave vector. both in the experiment
and in the calculations, but the calculated spin susceptibility is still
peaked \textbf{Q=(}$\pi,\pi),$ and the experiment also observes spin
fluctuations with the same wave vector. Also, the fact that FeBSC lack
strong Coulomb correlations\cite{Anis,Tom}  speaks against the former
alternative.

Recently, however, a new FeBSC, Sr$_{2}$VO$_{3}$FeAs, has been discovered
which seemingly violates this so far meticulously observed rule. The
calculated Fermi surface (FS)\cite{WEP} appears to be much more complex than
in the other investigated FeBSC, and there is no visual indication of any
quasinesting topology. Lee and Pickett\cite{WEP} argued that Sr$_{2}$VO$_{3}$%
FeAs represents \textquotedblleft a new paradigm for Fe-pnictide
superconductors\textquotedblright, and inferred that \textquotedblleft there
is no reason to expect an s$_{\pm}$ symmetry of superconducting order
parameter ($i.e.$ a different sign on the two FSs) in Sr$_{2}$VO$_{3}$FeAs.

\begin{figure}[tbp]
\includegraphics[width=0.95 \linewidth]{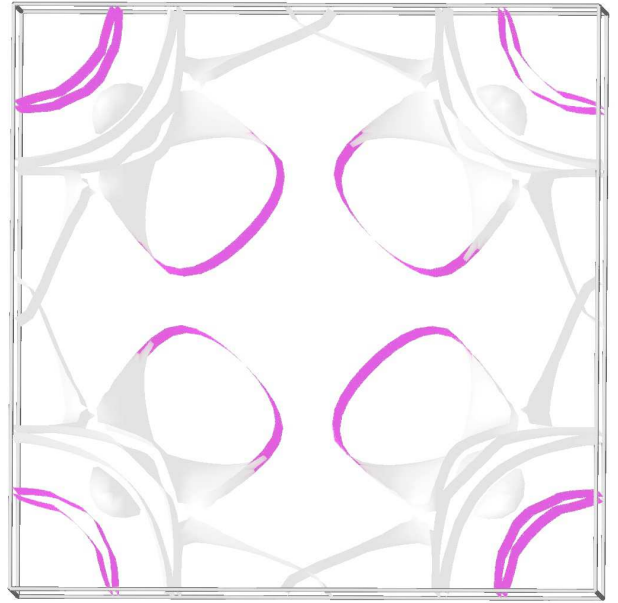}
\caption{The Fermi surfaces of Sr$_{2}$VO$_{3}$FeAs. The $\Gamma $ points are in the corners,
the M point in the center of the shown Brillouin zone. The colored (dark) portion are
the parts with the predominantly Fe character. The rest is predominantly V.
(color online)}
\label{FS}
\end{figure}
\begin{figure}[ptb]
\includegraphics[width=0.95 \linewidth]{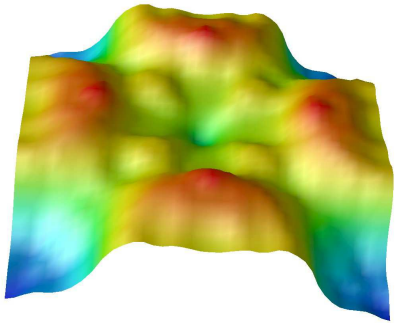}
\caption{The bare susceptibility (the real part) calculated with a constant
matrix element independently of the wave function character. The band
structure had been averaged over $k_z$ before the integration. The corners
of the plot correspond to $\mathbf{q}=(0,0)$, $(\protect\pi,0)$, $(0,\protect%
\pi)$, and $(\protect\pi,\protect\pi)$. The vertical scale is in arbitrary
units. (color online)}
\label{chi0}
\end{figure}

I have repeated the calculations of Lee and Pickett and have obtained the FS
that was similar to theirs\cite{note} (Fig. \ref{FS}). I have also verified
that the bare susceptibility \textit{without any account for the matrix
elements}%
\begin{equation}
\chi _{0}(\mathbf{q)}=-\sum_{\mathbf{k\alpha\beta }}\frac{f(\varepsilon _{\mathbf{%
k\alpha }})-f(\varepsilon _{\mathbf{k+q,\beta }})}{\varepsilon _{\mathbf{%
k\alpha }}-\varepsilon _{\mathbf{k+q,\beta }}+i\delta }
\end{equation}%
indeed does not have any peak at \textbf{Q=(}$\pi ,\pi )$ (Fig. \ref{chi0}).
In fact, it has a peak at an entirely different wave vector, $(\pi ,0.4\pi ),
$ as anticipated by Lee and Pickett. However, this does not take into
account the fact that the calculated Fermi surface is really a superposition
of two FS systems, one originating from the FeAs planes, and the other from
VO ones. While there is some hybridization between the two systems of bands
(at least along the XM direction; see Ref. \onlinecite{WEP} for details), as
well as a magnetic coupling and a magnetic moment on V, and maybe even
Coulomb correlation effects on V site, electrons derived from the Fe $d$%
-orbitals couple mostly with the spin fluctuations \textit{on the Fe sites}.
This is a simple consequence of the Hund's rule. With that in mind, I
colored the parts of the Fermi surface in Fig. \ref{FS} that have
predominantly Fe character.

\begin{figure}[ptb]
\includegraphics[width=0.95 \linewidth]{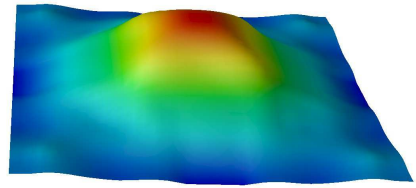} %
\includegraphics[width=0.95 \linewidth]{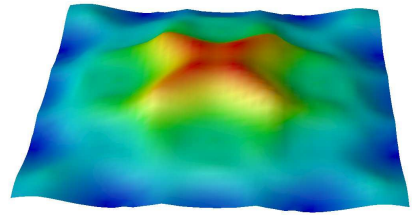}
\caption{The bare susceptibility calculated as in Fig.\protect\ref{chi0},
but with matrix elements taken as the product of the Fe weights for the
corresponding wave functions. The top panel shows the real part, the bottom
one the imaginary part. (color online)}
\label{chi}
\end{figure}
Imagine now that the \textit{unpainted} parts of the FS disappear. What
remains after this mental \textit{tour de force} closely resembles the
familiar FSs of other FeBSC. Taking into account the above argument
regarding the special role of the Fe spin fluctuations, we can rewrite Eq. 1
as 
\begin{equation}
\tilde{\chi}_{0}(\mathbf{q)}=-\sum_{\mathbf{k\alpha\beta}}\frac{f(\varepsilon _{%
\mathbf{k\alpha}})-f(\varepsilon_{\mathbf{k+q,\beta}})}{\varepsilon _{%
\mathbf{k\alpha}}-\varepsilon_{\mathbf{k+q,\beta}}+i\delta}A_{\mathbf{%
k\alpha}}A_{\mathbf{k+q,\beta}},
\end{equation}
where $A_{\mathbf{k\alpha}}$ is the relative weight of the Fe orbitals in
the $|\mathbf{k}\alpha\mathbf{>}$ wave function. The result (Fig. \ref{chi}%
), as expected, shows the same structure as for the other pnictides,
especially for the real part of susceptibility, which is the one relevant
for superconductivity.

I conclude that, unfortunately, Sr$_{2}$VO$_{3}$FeAs, despite being an
interesting and in many aspects unusual FeBSC, does not represent a new
paradigm, but rather falls into the same class as other pnictides. It is
also worth noting that while it has been established both experimentally\cite%
{Anis,Tom} and computationally\cite{Anis,Antoin} that the FeAs subsystem is
only weakly correlated, this had not been obvious \textit{a priori}, and it
is not obvious for the V-O subsystem in Sr$_{2}$VO$_{3}$FeAs. Being
essentially in a vanadium oxide layer (and vanadium oxide is strongly
correlated in the bulk form), V in Sr$_{2}$VO$_{3}$FeAs may be subject to
strong Hubbard correlations that would remove V states from the Fermi level%
\cite{LDAU}. Thus, strictly speaking, the conclusion above should be
formulated as follows: \textit{even if} Sr$_{2}$VO$_{3}$FeAs is a weakly
correlated metal and the FS calculated within the density functional theory
is realistic, the fact that the overall topology seems on the first glance
to be different from other pnictides is misleading and the spin fluctuation
spectrum is likely to be rather similar.

At the end, let me briefly touched upon a separate, but equally (if not
more) interesting issue of the magnetic ground state and magnetic properties
of Sr$_{2}$VO$_{3}$FeAs within the density functional theory (DFT). It is
well know\cite{review} that DFT seriously overestimates the tendency to
magnetism in FeBSCs, so that the calculated ground state appears strongly
antiferromagnetic even in the materials that sho no long range magnetic
order (phosphates, selenide). This is routinely ascribed to the mean-filead
character of DFT. However, it is of course interesting to see what is the
(magnetic) ground state in the mean filed, even when in real life the ground
state is paramagnetic. For all  FeBSCs studied so far the antiferromagnetic
stripe magnetic structure is by far the lowest in energy (energy gain of the
order of 200 meV per Fe compared to a nonmagnetic solution), while the
ferromagnetic structure is barely stable if at all.

\begin{figure}[ptb]
\includegraphics[width=0.95 \linewidth]{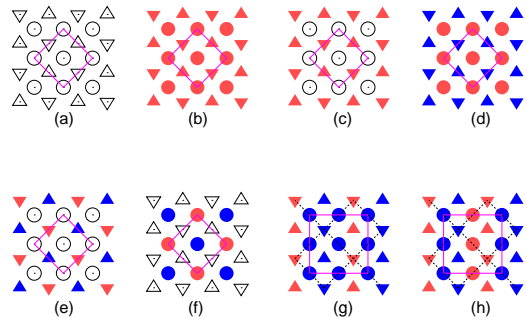}
\caption{Magnetic configurations used in the Table 1. Hollow symbols
indicate nonmagnetic atoms, blue (dark) spin-up moments, red (light)
spin-down moments. Circles are Fe atoms, upward and downward pointing
triangles are two V layers in the unit cell. The configurations are:
(a) NM:nonmagnetic, (b) FM:ferromagnetic, (c) half-FM, (d) FiM:
ferrimagnetic (Fe and V spins are antiparallel),
(e) V-AF: antiferromagnetically stacked FM V 
layers, nonmagnetic Fe (f) Fe-cb: checkerboard Fe planes, weakly ferromagnetic V planes,
(g) V-cb: checkerboard V planes, ferromagnetic Fe planes (h) V-cb combined with Fe stripes. Minimal crystallographic unit cell is
shown in each case, and in the last panel dashed lines connect V atoms in the
same layer (color online)}
\label{mag}
\end{figure}

Most likely, the DFT ground state of  FeBSCs is also antiferromagnetic
in-plane. However, even the nonmagetic unit cell contains 16 atoms, which
makes it extremely difficult to investigate the energy landscape for
possible antiferromagnetic pattern. Thus, it makes sense to study possible
ferro(ferri)magnetic solutions, in hope to extract at least some useful
information. This approach was adapted in Ref. \cite{Shein} (although these
authors do not present any nonmagnetic calculations, actually relevant for
superconductivity). They found a solution with a moment on V ($\sim 1.5$ $%
\mu _{B}),$ but not on Fe. Lee and Pickett found another, ferrimagnetic
solution, with opposite moments on V and Fe, the former being larger\cite{LP}%
. Using different starting configurations, I was able to converge to three
different ground states within the same symmetry, as shown in the Table, as well 
as to two lower-symmetry states, 
as illustrated in Fig. \ref{mag}(b,c,d): interlayer antiferromagnetic V sublattice,
where the V layers are ferromagnetic, and antiferromagnetically stacked,
while Fe is nonmagnetic, 
and Fe-checkerboard, where Fe forms a Neel plane 
and V in nonmagnetic. 
After that, I have calculated two configurations in the double (four
formula units) cell, which I feel are the most relevant because of the
superexchange interaction in the V layers: V-checkerboard with
nonmagnetic Fe, and V-checkerboard combined with the stripe order in the
Fe layers (Fig. \ref{mag})

A few observations are in place: (1) the state found in Ref. \cite{Shein} is
not the ground state even within that symmetry; (2) unlike all other FeBSCs,
FeAs planes can support a very stable ferromagnetic state; (3) the interaction
between V and Fe is ferromagnetic, that is, not of superexchange character,
(4) the
magnetic coupling between V and Fe is so weak that V does not induce any
magnetization on Fe, unless one already starts with a magnetic Fe; (5) It is more important,
from the total energy point of view, to have magnetic moment on V that on Fe --- a bit
surprising, given that V has a weaker Hund's rule coupling; (6) V
sublattice itself has a net antiferromagnetic interaction: if Fe is not
magnetic, V orders antiferromagnetically%
; 
(7) Unless some more exotic ground state will be discovered, the total
energy is minimized when V layers order in the Neel (checkerboard)
fashion, while Fe orders the same way as in other pnictides, forming
stripes;
(8)  most
importantly, a number of very different magnetic states are nearly degenerate in
energy. This last fact may be the key to the experimental fact that the
actual material is paramagnetic despite the fact that on the mean field
level it is more magnetic than other pnictides. This is an extremely
intriguing situation and the magnetism Sr$_{2}$VO$_{3}$FeAs deserves a more
elaborated experimental and theoretical study that is beyond the scope
of this paper.

\begin{table}
\caption{Properties of some stable magnetic solutions in the 
Generalized Gradient Approximation of the DFT. All energies are given
with respect to the nonmagnetic state. The magnetic states are described 
in Fig. \protect\ref{mag}. For the V-cb configuration I was able to 
converge to two different solution, high-spin and low-spin, with
essentially the same energies}
\begin{tabular}{|l|c|c|c|}
\hline
& $M_{Fe},$ $\mu _{B}$ & $M_{V},$ $\mu _{B}$ & $\Delta E.$ meV/Fe \\ 
\hline
FM & 2.0 & 1.4 & $-396$ \\ 
half-FM & 0.0 & 1.5 & $-381$ \\ 
FiM & 2.1 & -1.4 & $-387$\\
AFM-V& 0.1 &$\pm 1.4$&$ -385$\\
Fe-cb& $\pm$2.0 &0.2&$ -219$\\
V-cb&2.0& $\pm$1.2&-237\\
V-cb&0.1& $\pm$1.2&-232\\
V-cb + Fe-stripes&$\pm$2.2 &$\pm 1.2$&$ -409$\\
\hline
\end{tabular}
\end{table}

I thank W.E. Pickett and D. Parker for stimulating discussions related to
this work. I also acknowledge funding from the Office of Naval Research.

After this paper was accepted for publication, I became aware of another
band structure calculation\cite{China}. These author have considered the
``Shein-Ivanovskii'' half-FM states and two
antiferromagnetic states, with the checkerboard (Neel) and stripe ordering in
the Fe subslattice, and unspecified, presumably ferromagnetic, ordering
in the V subsystem. As clear from the above, neither in this states
represents an energy minimum even within the corresponding symmetry
group, therefore these authors arrived to an incorrect conclusion that
the lowest-energy magnetic state is characterized by Neel order in the
Fe subsystem.


\begin{thebibliography}{99}
\bibitem{note} I used the standard LAPW code as implemented in the WIEN2k
package with generalized gradient corrections. The orbitals weight used for
the rest of the calculations are the LAPW orbital projection inside the
corresponding muffin-tin spheres.
\bibitem{LDAU} Unfortunately, it is not clear at all whether the LDA+U
treatment would be appropriate for V in this fluctuating system even if V
itself is subject to strong Hubbard correlations.
\end{thebibliography}
\end{document}